\documentclass[aps,prl,twocolumn,superscriptaddress]{revtex4}

\usepackage{graphicx}

\begin{document}

\preprint{}

\title{Spin Susceptibility of Noncentrosymmetric Heavy-fermion Superconductor CeIrSi$_3$ under Pressure: $^{29}$Si-Knight Shift Study on Single Crystal}

\author{H. Mukuda}
\email[]{e-mail  address: mukuda@mp.es.osaka-u.ac.jp}
\author{T. Ohara}
\author{M. Yashima}
\author{Y. Kitaoka}
\affiliation{Graduate School of Engineering Science, Osaka University, Osaka 560-8531, Japan}

\author{R. Settai}
\author{Y. \={O}nuki}
\affiliation{Department of Physics, Graduate School of Science, Osaka University, Osaka 560-8531, Japan}

\author{K. M. Itoh}
\affiliation{Department of Applied Physics and Physico-Informatics, Keio University, Yokohama 223-8522, Japan}

\author{E. E. Haller}
\affiliation{University of California at Berkeley and Lawrence Berkeley National Laboratory, Berkeley, CA 94720, USA}

\date{\today}

\begin{abstract}
We report $^{29}$Si-NMR study on a single crystal of the heavy-fermion superconductor CeIrSi$_3$ without an inversion symmetry along the c-axis. The $^{29}$Si-Knight shift measurements under pressure have revealed that the spin susceptibility for the $ab$-plane decreases slightly below $T_c$, whereas along the $c$-axis it does not change at all.  The result can be accounted for by the spin susceptibility in the superconducting state being dominated by the strong antisymmetric (Rashba-type) spin-orbit interaction that originates from the absence of an inversion center along the $c$-axis and it being much larger than superconducting condensation energy. This is the first observation which exhibits an anisotropy of the spin susceptibility below $T_c$ in the noncentrosymmetric superconductor dominated by strong Rashba-type spin-orbit interaction. 
\end{abstract}


\maketitle

%%%%%%%%%%%%%%%%%%%%%%%%%%%%%%%%%  Introduction  %%%%%%%%%%%%%%%%%%%%%%%%%%%%
%\section{Introduction}

Recent discoveries of superconductors that lack inversion symmetry have aroused great interest because they possess quite unique properties\cite{Edelstein,Gorkov,Frigeri,FrigeriNJ,Samokhin,Hayashi,Fujimoto,Yanase,Bauer,Kimura,Sugitani,Settai,TateiwaCe113,MukudaCe113,Okuda,KimuraPRL2007,SettaiHc2,Togano,Yuan,Nishiyama}. Theoretically, the relationship between spatial symmetry and the spin state of Cooper-pairs is expected to be broken in those superconductors, which would make the parity of superconducting state mixed between even and odd parities due to the antisymmetric spin-orbit coupling (ASOC), thereby leading to a two-component order parameter composed of spin-singlet and spin-triplet Cooper pairing states~\cite{Edelstein,Gorkov,Frigeri,FrigeriNJ,Samokhin,Hayashi,Fujimoto,Yanase}. 
The heavy fermion (HF) superconductivity (SC) without inversion symmetry was first discovered in CePt$_3$Si~\cite{Bauer}. 
Subsequently, pressure ($P$)-induced SC was also observed in Ce-based HF antiferromagnets CeRhSi$_3$~\cite{Kimura}, CeIrSi$_3$~\cite{Sugitani}, and CeCoGe$_3$~\cite{Settai}, in which the inversion symmetry along the $c$-axis is broken.  
It is noteworthy that the SC transition temperature $T_{\rm c}$= 1.6 K for CeIrSi$_3$ under $P$= 2.6 GPa is relatively high among the Ce-based HF superconductors, which may be attributed to the strong coupling effect~\cite{TateiwaCe113} derived from the presence of strong antiferromagnetic (AFM) spin fluctuations~\cite{MukudaCe113}. 
The most remarkable feature in the SC state of CeIrSi$_3$ and CeRhSi$_3$ is that the upper critical field ($H_{c2}$) exceeds 30 T along the $c$-axis, which is much higher than a Pauli limiting field~\cite{KimuraPRL2007,Okuda,SettaiHc2}.
This kind of behavior has never been observed in other spin-singlet HF superconductors, such as CeCoIn$_5$, CeIrIn$_5$, CeCu$_2$Si$_2$, and so on. Actually, the ASOC of CeIrSi$_3$ has been estimated to be 1000 K from the Fermi surface splitting in LaIrSi$_3$~\cite{Okuda}, which is much larger than a superconducting condensation energy scaled to $k_{\rm B}T_c\sim$ 1.6 K. In this context, a novel SC state in noncentrosymmetric crystal structure would be expected for CeIrSi$_3$ and CeRhSi$_3$. 
%Recently, unconventional SC properties have been unraveled in the noncentrosymmetric compounds Li$_2$(Pt,Pd)$_3$B~\cite{Togano,Yuan,Nishiyama}. 
In general, the measurement of Knight shift by means of Nuclear Magnetic Resonance (NMR) can provide a clue to determine the spin state of the Cooper pair\cite{Tou}.   

In this Letter, we report on the $^{29}$Si-Knight shift measurement in the SC state of CeIrSi$_3$ under $P$=2.8 GPa. Its spin component slightly decreases below $T_c$ for the $ab$-plane, whereas it does not change along the $c$-axis. This result reveals that the spin susceptibility $\chi_{\rm s}(T)$ for noncentrosymmetric superconductors without inversion symmetry is dominated by the anisotropy of the {\it van-Vleck-like} spin susceptibility originating from the transition between the Fermi surfaces split by the ASOC, which makes it difficult to identify the admixture of the Cooper-pairing state between even and odd parity.

%%%%%%%%%%%%%%%%%%%%%%%%%%%%%%%   experimental    %%%%%%%%%%%%%%%%%%%%%%%%%%%%
%\section{Experimental}

A single crystal of CeIrSi$_3$ enriched by $^{29}$Si isotope was grown using a tetra-arc furnace, as described in the previous literatures~ \cite{Sugitani,Okuda}. A small piece of the single crystal, size of which is 1.3$\times$1.3$\times$5mm, was cut out from the ingot of a large single crystal by using the x-ray Raue method. Hydrostatic pressure was applied by utilizing a NiCrAl-BeCu piston-cylinder cell filled with an Si-based organic liquid as the pressure-transmitting medium. To calibrate the pressure at low temperatures, the shift in $T_{\rm c}$ of Sn metal under $P$ was monitored by its resistivity. 
$^{29}$Si-NMR measurements have been made at a magnetic field $H=$1.326 T.  In the experiment of SC state under pressure, the pressure cell  mounted on the mixing chamber was rotated within the plane including [110] and [001] directions, and fixed to the [110] or [001] directions by measuring the angle dependence of $^{29}$Si-NMR spectra. The single crystal was adjusted with an angle of less than $\pm4^\circ$ between the direction of $H$ and [110] or [001] directions, as described later.

%**************  Fig.1  *****************************************************
\begin{figure}[htbp]
\begin{center}
\includegraphics[width=0.8\linewidth]{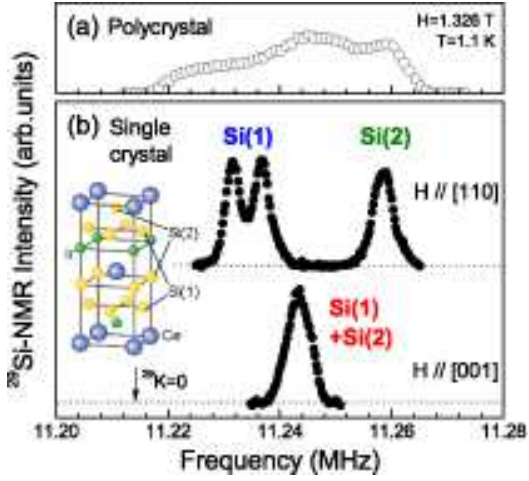}
\end{center}
\caption[]{\footnotesize (Color online) $^{29}$Si-NMR spectra  at $H=$ 1.326 T and $T=$ 1.1 K for (a) the polycrystal and (b) the single crystal of CeIrSi$_3$ under $P=$ 2.8 GPa. The NMR spectra at two inequivalent Si sites (denoted as Si(1) and Si(2)) are well separated in the single crystal. }
\label{fig:spectra_Poly_Single}
\end{figure}
%****************************************************************************

%%%%%%%%%%%%%%%%%%%%%%%%%%%%% results and discussions  %%%%%%%%%%%%%%%%%%%%%%%
%\section{Results and Discussions}

Figure~\ref{fig:spectra_Poly_Single}(b) shows $^{29}$Si-NMR spectra of the single crystal under $P=$ 2.8 GPa at $T=$ 1.1 K and $H=$1.326 T parallel to [110] and [001] directions. 
The linewidth of each spectrum is as narrow as $\sim$5 kHz, which is much narrower than that for the polycrystal shown in Fig.~\ref{fig:spectra_Poly_Single}(a), ensuring the high quality of the single crystal.
There are two crystallographically inequivalent Si sites denoted as Si(1) and Si(2), as illustrated in the inset of Fig.~\ref{fig:spectra_Poly_Single}(b). 
In $H\parallel[110]$, the NMR spectra arising from Si(1) and Si(2) are well separated, as seen in the upper part of Fig.~\ref{fig:spectra_Poly_Single}(b). Since the number of Si(1) is twice as much as that of Si(2), the spectrum with twin peaks at lower frequencies is assigned to Si(1) and hence the spectrum with a single peak to Si(2).
The twin peaks at Si(1) arise because $H$ is not exactly set along the [110] direction, but with approximately $\pm4^\circ$ from it. Note that the hyperfine field at Si(1), $A_{\rm hf}^{[110]}(1)$ and hence Knight shift, $K(1)^{[110]}$ is highly anisotropic within the $ab$-plane, originating from the pseudo-dipole field from $3p$ orbitals.
By contrast, the single peak at Si(2) reveals that an anisotropy in the Knight shift $K(2)^{[110]}$ within the $ab$-plane is much smaller than that in $K(1)^{[110]}$. 
When $H$ is applied to the [001] direction, the spectra at Si(1) and Si(2) overlap because the Knight shift ($K^{[001]}$) at both sites is almost the same.

%**************  Fig.2  *****************************************************
\begin{figure}[htbp]
\begin{center}
\includegraphics[width=0.85\linewidth]{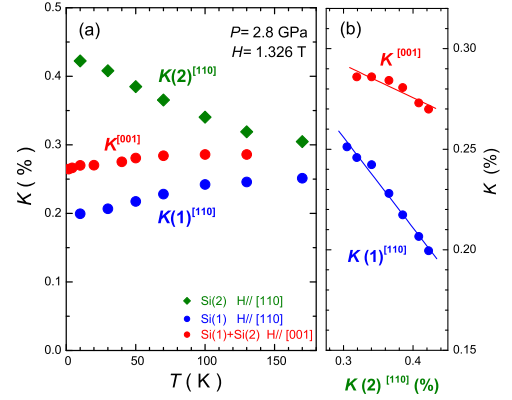}
\end{center}
\caption[]{\footnotesize(Color online) (a) $T$ dependence of the $^{29}$Si-Knight shift in the normal state under $P=$ 2.8 GPa. (b) Plots of $K(1)^{[110]}$ and $K^{[001]}$ versus $K(2)^{[110]}$ as the implicit parameter of $T$. Slopes in this plot give rise to the ratios of the hyperfine coupling constants, $A_{\rm hf}^{[110]}(1)/A_{\rm hf}^{[110]}(2)\approx-0.45$ and $A_{\rm hf}^{[001]}(1)/A_{\rm hf}^{[110]}(2)\approx-0.16$.}
\label{fig:Knightshift}
\end{figure}
%****************************************************************************

In general, Knight shift is given by,
\[
K(T) = K_{\rm s}(T) + K_{\rm orb} + K_{\rm dia}
\]
where $K_{\rm s}(T)$ and $K_{\rm orb}$ are the spin and orbital part of Knight shift, and $K_{\rm dia}$ is the SC diamagnetic shift.  
Note that $K_{\rm s}(T)=A_{\rm hf}\chi_{\rm s}(T)$ depends on $T$, where $A_{\rm hf}$ and $\chi_{\rm s}(T)$ are hyperfine coupling constant and spin susceptibility. $K_{\rm orb}$ is related to the {\it usual} van Vleck susceptibilities and hence does not depend on $T$. 
Figure~\ref{fig:Knightshift}(a) shows the temperature ($T$) dependencies of $K(1)^{\rm [110]}$, $K(2)^{\rm [110]}$ and $K^{\rm [001]}$ in the normal state up to 170 K under $P=$ 2.8 GPa. 
The $K(2)^{[110]}$ increases upon cooling, which is scaled to the bulk susceptibility $\chi(T)$ at ambient pressure~($P=0$)~\cite{Okuda}.  In contrast, $K(1)^{[110]}$ and $K^{[001]}$ decrease upon cooling because $A_{\rm hf}$s are negative. 
Note that $K(1)^{[110]}$, which was estimated from an average value of shifts at the twin peaks in the spectra of Si(1) in Fig.~\ref{fig:spectra_Poly_Single}(b), is in accord with the shift at the field along the $[110]$ direction.  
Because of the lack of $\chi(T)$ data under $P$, we could not make a plot of $K$~versus~$\chi$ to deduce $A_{\rm hf}$ and $K_{\rm orb}$.
However, as shown in Fig.~\ref{fig:Knightshift}(b), respective plots of $K(1)^{[110]}$ and $K^{[001]}$ versus $K(2)^{[110]}$ as the implicit parameter of $T$ give rise to the ratios of $A_{\rm hf}(1)^{[110]}/A_{\rm hf}(2)^{[110]}\approx-0.45$ and $A_{\rm hf}^{[001]}/A_{\rm hf}(2)^{[110]}\approx-0.16$, from their slopes of tentative linear lines that were indicated in the figure. 

%**************  Fig.3  *****************************************************
\begin{figure}[htbp]
\begin{center}
\includegraphics[width=0.8\linewidth]{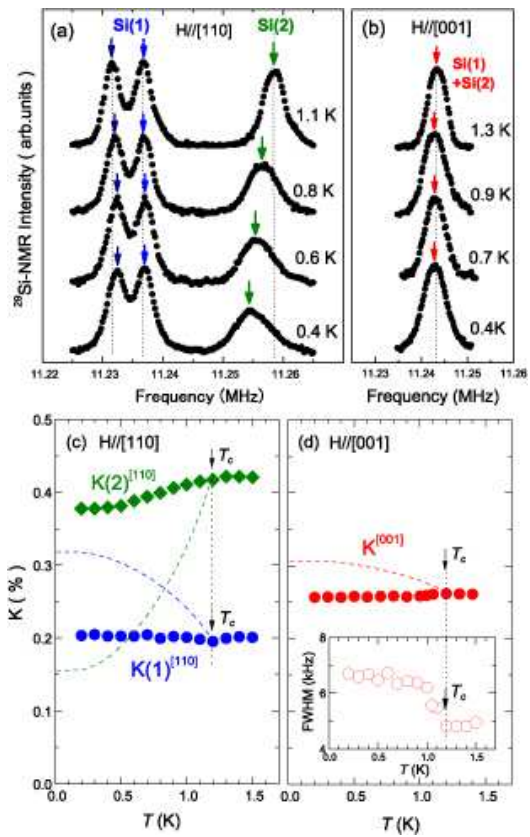}
\end{center}
\caption[]{\footnotesize(Color online) $T$ dependence of the $^{29}$Si-NMR spectra at $H=1.326$ T parallel to (a) [110] and (b) [001] directions below 1.3 K. The $T$ dependencies of $K(1)^{[110]}$ and $K(2)^{[110]}$ along (c) [110]  and $K^{[001]}$ along (d) [001] directions. The inset of (d) is the $T$ dependence of full-width at half-maximum (FWHM) of the spectrum for $H\parallel [001]$, which enables us to determine $T_c(H)=1.2$ K. Broken curves are calculations for a case of $\chi_{\rm s}(T)\to 0$ at low $T$ limit, assuming the residual DOS of 37\% at the $E_F$ deduced from the $1/T_1$ measurement on the polycrystal~\cite{MukudaCe113}.}
\label{fig:Knightshift_SCstate}
\end{figure}
%****************************************************************************

Next, we deal with the Knight shift measurements in the SC state. Figures \ref{fig:Knightshift_SCstate}(a-d) indicate the $T$ dependencies of $^{29}$Si-NMR spectra and the Knight shifts below 1.5 K under $P=$~2.8~GPa. 
The $T_c(H)=1.2$ K in the present experiment at $P=$2.8~GPa was determined from the result that the inhomogeneity of the field in the SC mixed state increases the NMR spectral width below 1.2 K as presented in the inset of Fig.~\ref{fig:Knightshift_SCstate}(d). 
In $H\parallel~[110]$, $K(2)^{[110]}$ decreases below $T_c$, pointing to the decrease of $\chi_{\rm s}(T)$. 
Owing to the opposite sign of $A_{\rm hf}$, $K(1)^{[110]}$ slightly increases. Using the relations of $K(2)^{[110]}=A_{\rm hf}(2)^{[110]}\chi_{\rm s}(T)$+$K_{\rm orb}(2)$+$K_{\rm dia}$ and $K(1)^{[110]}=A_{\rm hf}(1)^{[110]}\chi_{\rm s}(T)$+$K_{\rm orb}(1)$+$K_{\rm dia}$ where $K_{\rm dia}$ is identical at Si(1) and Si(2); and $A_{\rm hf}(1)^{[110]}/A_{\rm hf}(2)^{[110]}\approx-0.45$, $K_{\rm dia}$ was evaluated to be $\sim$0.006\% ($H_{dia}\sim$0.78 Oe in the field scale). 
Since an estimation of $K_{\rm orb}$ at $P=2.8$ GPa is difficult due to that a $K$-$\chi$ plot is not available, $K_{\rm orb}(2)=0$ at $P=2.8$ GPa is assumed as well as at $P=0$~\cite{Site}. 
A broken curve in Fig.~\ref{fig:Knightshift_SCstate}(c) is a calculated  curve for a case where $\chi_{\rm s}(T)$ would be reduced below $T_c(H)$ if a SC had a line-node gap with 2$\Delta/k_{\rm B}T_c\sim6$ and the residual DOS of 37\% at the Fermi level($E_F$), which was experimentally deduced from the $1/T_1$ measurement on the polycrystal~\cite{MukudaCe113}. 
Likewise, by assuming the same model and by using the ratios of $A_{\rm hf}(1)^{[110]}/A_{\rm hf}(2)^{[110]}\approx-0.45$ and $A_{\rm hf}^{[001]}/A_{\rm hf}(2)^{[110]}\approx-0.16$ obtained experimentally, the increasing events below $T_c$ would be expected in $K_{\rm s}(1)^{[110]}$ and $K_{\rm s}^{[001]}$, as shown by the broken curves in the figure.
In $H\parallel [001]$, $K^{[001]}$ should be increased if $\chi_{\rm s}(T)^{[001]}$ were reduced below $T_c(H)$, since $A_{\rm hf}^{[001]}$ is negative, as indicated in Fig.~\ref{fig:Knightshift_SCstate}(d). 
However, as decreasing $T$ below $T_c(H)=1.2$ K, $K^{[001]}$ slightly decreases by 0.006\%, which is comparable to $K_{\rm dia}$ along the [110] direction. If we assume that $K_{\rm dia}$ for $H\parallel [001]$ is nearly the same as for $H\parallel [110]$,  it might be expected that $K_{\rm s}^{[001]}$ does not change below $T_c$. 

%**************  Fig.4 *****************************************************
\begin{figure}[htbp]
\begin{center}
\includegraphics[width=0.9\linewidth]{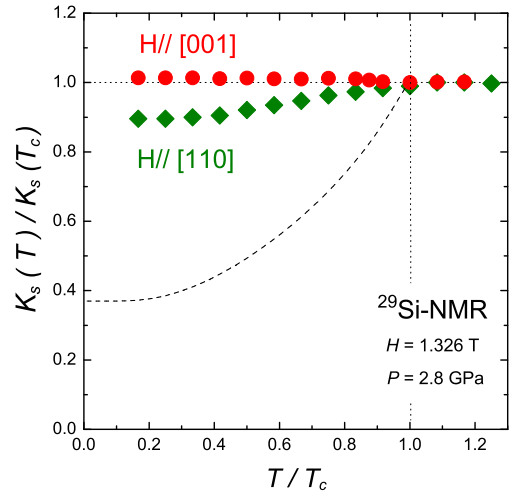}
\end{center}
\caption[]{\footnotesize(Color online) The $T$ dependences of the plots of $K_{\rm s}(T)/K_{\rm s}(T_c)$ versus $T/T_c$ for $H\parallel [110]$ and $[001]$ in the SC state. A broken curve is a calculation for a case of $\chi_{\rm s}(T)\to 0$ at low $T$ limit, assuming the residual DOS of 37\% at the $E_F$\cite{MukudaCe113}(See text). }
\label{fig:Knightshift_summary}
\end{figure}
%****************************************************************************

Figure \ref{fig:Knightshift_summary} shows the plots of $K_{\rm s}(T)/K_{\rm s}(T_c)$ versus $T/T_c$ for $H\parallel [110]$ and $[001]$. 
It is remarkable that $(K_{\rm s}(T)/K_{\rm s}(T_c))^{[110]}$ slightly decreases, whereas $(K_{\rm s}(T)/K_{\rm s}(T_c))^{[001]}$ does not decrease at all. Note that when {\it usual} superconductors with an inversion center are in a spin-singlet regime, $\chi_{\rm s}(T)$ decreases to zero generally {\it regardless of crystal directions}, and when those are in a spin-triplet regime, $\chi_{\rm s}(T)$ for $H\perp {\bf d}-vector$ stays constant, but that for $H|| {\bf d}-vector$ decreases to zero. 
It is known that the $K_{\rm s}(T)$ in some spin-singlet SC remains finite at low $T$ limit in association with an impurity and/or defects existing inevitably\cite{Curro}. 
As a matter of fact, the present result cannot be reproduced by the simulation for a case of $\chi_{\rm s}(T)\to 0$ at low $T$ limit, even though taking into account the residual DOS of 37\% at the $E_F$ due to the impurity effect\cite{MukudaCe113}, as shown by a broken curve in Fig. \ref{fig:Knightshift_summary}. 
In this context, it is anticipated that the present results for CeIrSi$_3$ with no inversion symmetry differ from these behaviours in the centrosymmetric superconductors. 

In inversion-symmetry-broken systems, it has been theoretically shown that $\chi_{\rm s}(T)$  consists of the Pauli term $\chi^{\rm Pauli}(T)$ and the van-Vleck-like term $\chi^{\rm VV}(T)$~\cite{Fujimoto}; The former originates from the Pauli paramagnetism due to the spin distribution on the intrabands. On the other hand, the latter originates from the transition between the  Fermi surfaces split by the strong ASOC in the noncentrosymmetric compounds, which differs in origin from the {\it usual} $T$-independent van Vleck susceptibility stemming from the transition between the different orbitals. 
A calculation assuming the simple spherical Fermi surface reveals that $\chi_{\rm s}^{\rm SC}/\chi_{\rm s}^{\rm N} \to 1$ for $H\parallel[001]$ and $\chi_{\rm s}^{\rm SC}/\chi_{\rm s}^{\rm N} \to 1/2$ for $H\parallel[110]$ well below $T_c$, {\it irrespective of the pairing symmetry}~\cite{FrigeriNJ}, provided that the Rashba-type ASOC is sufficiently larger than the SC gap~\cite{Okuda}. 
The experimental result is consistent in the case for $H\parallel[001]$, however, $\chi_{\rm s}^{\rm SC}/\chi_{\rm s}^{\rm N}\sim 0.9$ at $T\to 0$ for $H\parallel [110]$ is not consistent with the prediction based on this model. 
Fujimoto pointed out theoretically that the $\chi^{\rm VV}(T)/\chi_{\rm s}(T)$ in the SC state is intimately affected by electronic structures and electron correlation effects~\cite{Fujimoto}: In this case, the value of $\chi_{\rm s}^{\rm SC}/\chi_{\rm s}^{\rm N}$ at low $T$ limit corresponds to a fraction of $\chi^{\rm VV}(T)$ contributing to $\chi_{\rm s}(T)$. Consequently, the results of no change in $\chi_{\rm s}(T)$ for $H\parallel [001]$ across $T_c$, but its slight decrease of $\chi_{\rm s}(T)$ for $H\parallel[110]$ below $T_c$ suggest the anisotropy of $\chi^{\rm VV}(T)$, dominated by the strong ASOC. This finding may be a characteristic feature of the spin susceptibility in the noncentrosymmetric superconductor CeIrSi$_3$ with the Rashba-type ASOC, which yields a large contribution of $\chi^{\rm VV}(T)$ along the $c$-axis. 

This result also gives an indication for the large anisotropy of a Pauli limiting field ($H_P$) between $H\parallel[001]$ and $H\parallel[110]$ at $T$=0, which is roughly estimated through a comparison of a superconducting condensation energy and a spin polarization energy given by $H_P=k_{\rm B}T_c/\mu_B\sqrt{1-(\chi^{\rm SC}/\chi^{\rm N})}$~\cite{FrigeriNJ}. This fact gives a clue to understand  why $H_{c2}$ is much larger along the $c$-axis than along the $ab$-plane in this compound. 
%nevertheless a Pauli limiting field at $T$=0 for the $c$-axis and the $ab$-plane is evaluated to be roughly compatible from  a comparison of a superconducting condensation energy and a spin polarization energy given by $H_P=k_{\rm B}T_c/\mu_B\sqrt{1-(\chi^{\rm SC}/\chi^{\rm N})}$~\cite{FrigeriNJ}. 
However, it should be noted that one possible reason why $H_{c2}(0)$ in a narrow $P$ range around $P=$2.6 GPa is exceptionally enhanced~\cite{Settai} may be attributed to strong AFM spin fluctuations at magnetic quantum criticality~\cite{Tada_Hc2} or other unknown critical behaviors.
Although the decrease in  $\chi_{\rm s}(T)$ for $H\parallel [110]$ below $T_c$ evidences the Cooper-pair formation, it seems to be  difficult to argue which types of Cooper-pairing states, that are theoretically predicted in noncentrosymmetric superconductors, take place: a spin-singlet state, a spin-triplet one, or an admixture of those states~\cite{Edelstein,Gorkov,Frigeri,FrigeriNJ,Samokhin,Hayashi,Fujimoto,Yanase} for the case that the strong ASOC dominates the spin susceptibility.  

Finally, we comment on the Knight-shift measurements in the HF superconductor CePt$_3$Si without an inversion center along the $c$-axis~\cite{Bauer}: No change of $^{29}$Si-Knight shift for $H$ parallel and perpendicular to the $c$-axis was observed  in the samples either with $T_c=$0.75 K~\cite{Yogi2006} or $T_c=$0.46 K that is inherent to CePt$_3$Si~\cite{Takeuchi,MukudaCe131,TcCePt3Si}. An anisotropy in the Knight shift was not resolved in CePt$_3$Si, if any, because the linewidth of $^{29}$Si-NMR spectra was one order larger than in the single crystal of CeIrSi$_3$ due to the coexistence of AFM order and  the SC in CePt$_3$Si at $P$=0.  

%\section{Conclusion}

In conclusion, the extensive measurements of $^{29}$Si Knight shift for the noncentrosymmetric CeIrSi$_3$ under pressure have revealed that the spin component of the Knight shift $K_{\rm s}^{[110]}$ in the $ab$-plane slightly decreases, whereas $K_{\rm s}^{[001]}$ along the $c$-axis is unchanged across $T_c$. This is considered because the spin susceptibility in the SC state is dominated by the van Vleck-like spin susceptibility in association with the strong ASOC caused by the absence of an inversion center as pointed out theoretically. This is the first observation which exhibits an anisotropy of the spin susceptibility below $T_c$ in the noncentrosymmetric superconductor dominated by the strong Rashba-type spin-orbit interaction. We also comment that the Knight shift measurement does not give a direct evidence for the admixture of the Cooper-pairing state between even and odd parity in the case that the spin susceptibility is dominated by strong spin-orbit interaction. 

%\section*{Acknowledgements}

We would like to thank S. Fujimoto for his valuable comments. 
This work was supported by a Grant-in-Aid for Specially Promoted Research (20001004) and by Global COE Program (Core Research and Engineering of Advanced Materials-Interdisciplinary Education Center for Materials Science) from the Ministry of Education, Culture, Sports, Science and Technology (MEXT), Japan.

%::::::::::::::::bibliography::::::::::::::::::::::::::::::::::::::::::::::::
%:::::::::::::::::::::::::::::::::::::::::::::::::::::::


\begin{thebibliography}{99}
\bibitem{Site} In the previous work using the polycrystal~\cite{MukudaCe113}, the NMR peak with a largest Knight shift was tentatively assigned to Si(1). However, in the present work on the single crystal, it was assigned to Si(2). Accordingly, $K_{\rm orb}$ at Si(2) is evaluated to be almost zero at $P=0$.
\bibitem{TcCePt3Si} The $T_c$ of the high-quality single crystal was reported to be 0.46 K~\cite{Takeuchi}. The NMR study suggested the presence of two domains in CePt$_3$Si; the large domains without disorder and small ones with disorder set in SC at $T_c=$ 0.46 K and $T_c=$ 0.75 K, respectively\cite{MukudaCe131}.
\end{thebibliography}
\end{document}